\documentclass[aip,cha,preprint]{revtex4-1}
\usepackage{amsmath}
\usepackage{amssymb}
\usepackage{graphicx}
\usepackage{dcolumn}
\usepackage{bm}
\usepackage{epsfig}
\usepackage{natbib}
\usepackage{psfrag}

\newcommand {\e} {\varepsilon}

\begin{document}
\title{
Spreading of energy in the Ding-Dong Model}

\author{S. Roy}
\affiliation{Department of Physics and Astronomy, University of Potsdam, 
Karl-Liebknecht-Str. 24/25, 14476 Potsdam, Germany}
\affiliation{Department of Physics,
Indian Institute of Technology, Kanpur, India}
\author{A. Pikovsky}
\affiliation{Department  of Physics and Astronomy, University of Potsdam, 
Karl-Liebknecht-Str. 24/25, 14476 Potsdam, Germany}

\date{\today}

\begin{abstract}
We study properties of energy spreading in a lattice of elastically colliding harmonic oscillators (Ding-Dong model). 
We demonstrate that in the regular lattice the spreading from a localized initial state is mediated by compactons and chaotic
breathers. In a disordered lattice the compactons do not exist, and the spreading eventually stops, resulting in a 
finite configuration with 
a few chaotic spots. 
\end{abstract}

\maketitle

\begin{quotation}
In a linear medium energy propagates in form of waves, e.g. as phonons in a chain of linearly interacting particles.
Disorder leads to Anderson localization which blocks the spreading. In a nonlinear medium, 
chaos may appear, so the Anderson 
localization is destroyed and a weak subdiffusive spreading of energy is observed. Still, it is not clear how the final stage of
this very slow process looks like. There is a class of systems where the interaction between particles is purely nonlinear, here phonons 
do not exist and all spreading mechanisms are essentially nonlinear. In such ordered strongly nonlinear lattices, 
nonlinear waves -- compactons -- may transport energy to large distances, while in disordered case one observes a slow subdiffusive
energy spreading due to chaos. The Ding-Dong model that we study in this paper is a simple 
although singular
realization of a strongly nonlinear lattice,
where linear oscillators interact due to elastic collisions. Numerical simulations of this system are
very effective, and allow us to characterize the properties of compactons and energy spreading in disordered lattices at very large times. 
Our conclusion is that in the disordered case the spreading eventually stops, resulting in a few chaotic spots where 3-4
neighboring particles collide in an irregular manner.
\end{quotation}

\section{Introduction}
Dynamics of nonlinear lattices is one of the central topics 
of nonlinear science. Since the pioneering works of Fermi, Pasta, and Ulam (see Refs.~\cite{Gallavotti-08,Chaos-fpu-05}) 
enormous progress has been achieved in understanding 
of different aspects of interesting dynamical properties of such systems, in particular of 
solitary waves (solitons and compactons), thermalization, transition to chaos.   Very popular in these studies are 
models of elastically colliding particles. Indeed, such systems have been proved to be tractable
in deriving foundations of statistical mechanics from dynamical equations~\cite{Sinai-Chernov-87}. In the context of 
one-dimensional lattices, in papers~\cite{Casati84,Casati-86} a so-called ``Ding-a-Ling'' model has been introduced, 
consisting of
an alternate sequence of equal mass, hard point, free and bounded by harmonic forces elastically colliding particles.
For further studies of this model see Refs.~\cite{Posch-Hoover-98,Gawronski-Kulakovski-02}. 

In this paper we study a more simple
Ding-Dong model, introduced by Robnik and Prosen~\cite{Prosen-Robnik-92}. This is a lattice of colliding harmonic oscillators. Our main 
interest is in the general features of energy spreading in such lattices: how an initially localized perturbation (wave packet) spreads. We will 
see that this process can be regarded as an interplay of compactons (strongly localized solitary waves) and chaos. 

The problem of energy spreading in nonlinear lattices has attracted large interest recently in the context of disordered systems. 
It is known, that linear disordered lattices demonstrate Anderson localization, i.e. the spreading is effectively blocked by 
disorder. Nonlinearity typically destroys the localization, leading to a very slow 
energy spreading~\cite{Shepelyansky-93,%
Molina-98,%
Pikovsky-Shepelyansky-08,Garcia-Mata-Shepelyansky-09,%
Flach-Krimer-Skokos-09,Skokos_etal-09,%
Mulansky-Ahnert-Pikovsky-Shepelyansky-09,%
Skokos-Flach-10,Flach-10,Laptyeva-etal-10,%
Mulansky-Pikovsky-10,Johansson-Kopidakis-Aubry-10,Mulansky-Ahnert-Pikovsky-11}. One of the still unsolved issues here is whether the spreading 
occurs indefinitely (although slowly) or is eventually stopped. The last possibility might be reasonable, as in course of spreading the energy
density decreases and the system becomes closer and closer to the linear one.
In general nonlinear disordered lattices the answer to this dilemma is very difficult and relies on a heavy numerical integration
(cf. Refs.~\cite{Johansson-Kopidakis-Aubry-10,Mulansky-Ahnert-Pikovsky-11}). 
In this paper we attack this problem for the Ding-Dong model, where numerical simulations are 
much more effective.

The paper is organized as follows. First we introduce the Ding-Dong model. Then we discuss compactons and chaotic 
breathers. Spreading of energy is studied next for homogeneous and disordered lattices. We conclude with a discussion.

\section{Ding-Dong Model}

The Ding-Dong model have been first formulated and studied by T. Prosen and M. Robnik~\cite{Prosen-Robnik-92}. 
It is a chain of harmonic oscillators with a Hamiltonian
\begin{equation}
 H=\sum_k\left(\frac{p_k^2}{2}+\frac{q_k^2}{2}\right)\;.
\label{eq:ddbase}
\end{equation}
The oscillators are aligned along a line with a spacing distance $1$, so that if $q_k=1+q_{k+1}$, an elastic collision
between the oscillators $k$ and $k+1$ occurs, at which they exchange their momenta $p_k\to p_{k+1}$, $p_{k+1}\to p_k$.
Together with the total energy, the energy of the center of mass motion $E_{cm}=\frac{1}{2}((\sum_k p_k)^2+(\sum_k q_k)^2)$
is conserved.

The simplicity of the dynamics allows one a very efficient integration strategy of this nonlinear oscillator chain, at which
the collision times are determined explicitly by using standard inverse trigonometric functions only. This allows 
one to proceed to very large times
with the accuracy of the double precision numerical arithmetic. Contrary to previous studies of the model (\ref{eq:ddbase}), 
where the focus of the 
interest was on the heat conductivity properties of the 
lattice attached to thermostats~\cite{Prosen-Robnik-92,Sano-00,Sano-Kitahara-01,Sano-06}, 
we 
are interested in the \textit{energy spreading} problem. We assume that initially only a small localized
 part of the lattice is excited, while all other oscillators are at rest, and characterize the spreading of the energy from such 
an initial configuration. We will see that 
the main ingredients of the dynamics are compactons and chaotic breathers, which we discuss in the next section.

\section{Compactons and chaotic states}

\subsection{Examples of compactons}
The notion of compactons as of solitary waves with a compact support in nonlinear systems has been introduced by 
Ph. Rosenau~\cite{Rosenau-Hyman-93,Rosenau-94,Rosenau-06}. Such compact solutions exist in strongly 
nonlinear partial differential equations. In strongly nonlinear lattices the corresponding traveling waves and breathers are
not strictly compact, but have superexponentially fast decaying 
tails~\cite{Rosenau-Schochet-07,Rosenau-Pikovsky-05,Pikovsky-Rosenau-06,Ahnert-Pikovsky-08,Ahnert-Pikovsky-09}.
Remarkably, in the Ding-Dong model (\ref{eq:ddbase}) 
strictly compact traveling waves exist. One family of such waves has been
found by T. Prosen and M. Robnik~\cite{Prosen-Robnik-92}, we call these one-particle compactons, as 
these are solutions where at some moment of 
time just one oscillator in the lattice is excited (i.e. has non-zero energy). 
The family is determined by the coordinate and the momentum at the excited site: 
$q=0,p=[\cos(\pi/(2(n+1)))]^{-1}$. 

\begin{figure}[!hbt]
\centering
\includegraphics[width=\textwidth]{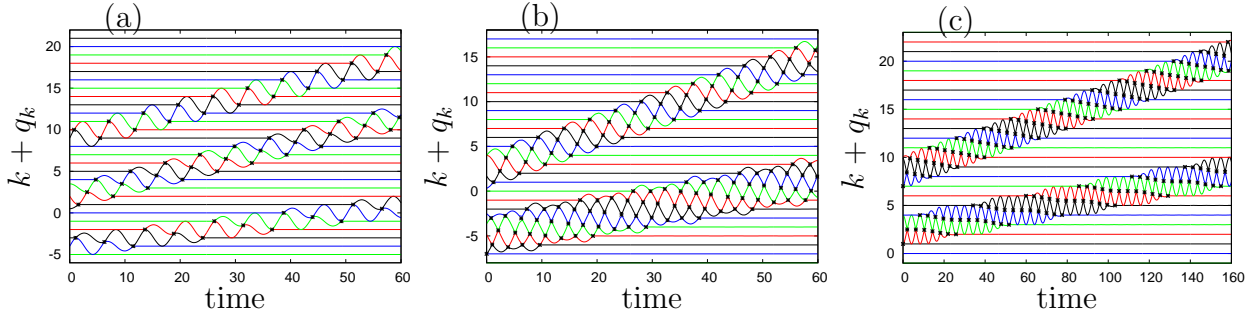}
\caption{In (a) we show two Robnik-Prosen compactons (upper and lower pulses) with $n=1,2$, together with the only one 
two-particle compacton, for which we have an analytic expression $-q_{-1}=q_1=0.5,\; p_1=-1,\; q_0=p_{-1}=p_1=0$. 
All other multi-particle compactons shown in (b),(c) have been found numerically. Lines are trajectories of the particles, 
markers depict collisions.
}
\label{fig:cmp1}
\end{figure}

We have found several other compactons, which are many-particle pulses, i.e. at each moment of time at 
least several oscillators are 
excited (only for one of these waves we have found an analytical formula, all other are obtained numerically).  
These compactons, together with two representatives of the Robnik-Prosen family,  are presented in Fig.~\ref{fig:cmp1}.
 
\subsection{Stability of compactons}
\begin{figure}[!hbt]
\centering
\includegraphics[width=0.9\textwidth]{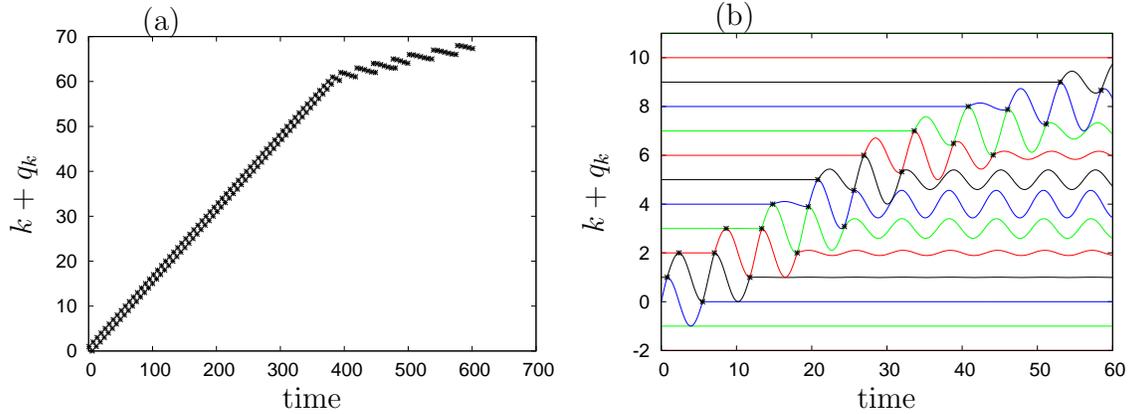}
\caption{Evolution of the perturbed compacton (cf. top pulse in Fig.~\ref{fig:cmp1}(a)). (a): relatively large negative perturbation 
$\e_0=0.016$, (b): small positive perturbation $\e_0=-10^{-10}$.}
\label{fig:cmp_st}
\end{figure}

One-particle compactons possess a remarkable one-side stability. We illustrate this in Fig.~\ref{fig:cmp_st}, where
we show the evolution for a slightly disturbed compacton with $n=1$ (it has unit energy) from the Robnik-Prosen family. For initial
energies slightly less than $1$ the compacton survives for a very long time, 
even if the perturbation is relatively large (Fig.~\ref{fig:cmp_st}(a)), 
while for energies slightly larger than
$E=1$ its life time is very short (Fig.~\ref{fig:cmp_st}(b)).

To understand the one-side stability, we apply a perturbative approach. Suppose that the initial energy of particle $0$ 
is $E_0=1-\e_0$, 
$\e_0\ll 1$. Then at the next collision the particle $1$ will get velocity $V_1=\sqrt{2E_0-1}$. After some time $t$, defined from the
equation $V_1\sin t-\cos t+1=0$, particles $0$ and $1$ will collide again, and after this collision the particle $0$ will have energy
$\tilde E_0=E_0\cos^2t $. This energy is ``lost'' due to the perturbation. Determining $t$ to the leading order in $\e_0$, we find
$\tilde E_0=\e_0^2$. Thus, particle $1$ will have energy $E_1=1-\e_0-\e_0^2$, and the whole cycle repeats.
 Summarizing, we obtain the following 
equation for the
energy losses:
\begin{equation}
 \e_{n+1}=\e_n+\e_n^2\;.
\end{equation}
Approximating the evolution as a continuous one by replacing $\e_{n+1}-\e_n\to\frac{d\e}{dn}$ we obtain that the losses grow as
\begin{equation}
 \e_n=\frac{\e_0}{1-n\e_0}\;.
\end{equation}
Thus, the life time of the compacton (in propagation sites) is $\sim (\e_0)^{-1}$. 

The life time for the positive 
perturbations ($\e_0<0$) is extremely small, so that we could not construct a perturbation theory and
find an expression for it. One can see from 
Fig.~\ref{fig:cmp_st}(b) that for  $\e_0=-10^{-10}$ the life time of a compacton is around 10.

\begin{figure}[!hbt]
\centering
\includegraphics[width=0.9\textwidth]{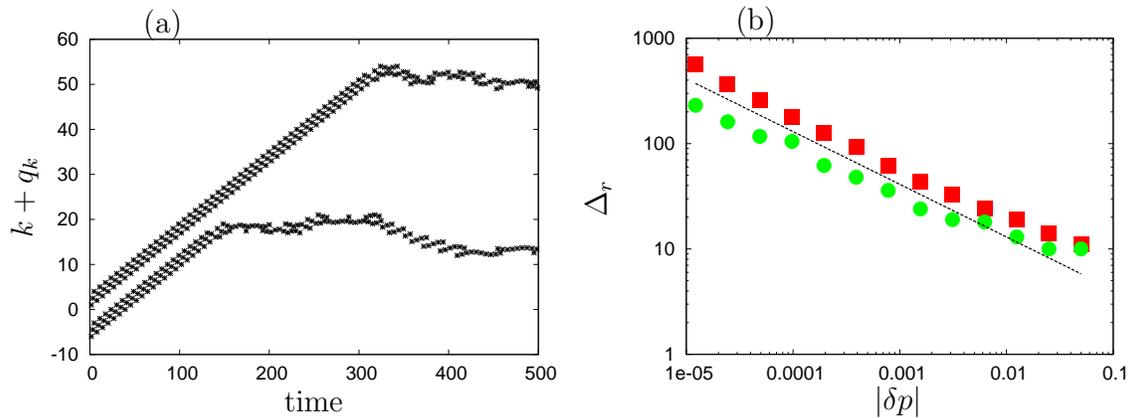}
\caption{(a) Evolutions of a perturbed two-site compacton (middle pulse in Fig.~1(a)) 
for $\delta p=-10^{-3}$ (upper case) and $\delta p=10^{-3}$ (bottom case).
(b) Maximal propagation of a compacton in dependence on $\delta p$ (squares negative $\delta p$, circles 
positive $\delta p$; the line has slope $-0.5$).}
\label{fig:cmp_st_3}
\end{figure}

The simplest multi-site compacton shown in Fig.~\ref{fig:cmp1}(a) is less stable to perturbations than the one-particle 
compacton, but its stability is relatively symmetric to the sign of the perturbation. We illustrate this in Fig.~\ref{fig:cmp_st_3}. 
There we perturb the initial state of the compacton as $q(-1)=-q(1)=-0.5$, $p(0)=-1+\delta p$, and all other initial momenta 
and positions are zero. While the case of positive $\delta p$ is slightly more unstable than that of negative perturbations, 
the maximal propagation length of the compacton $\Delta_r$ (defined as the position of the maximal site having non-zero energy)
in both cases scales as $\sim (|\delta p|)^{-1/2}$.

\subsection{Chaotic states}

\begin{figure}[!hbt]
\centering
\includegraphics[width=0.45\textwidth]{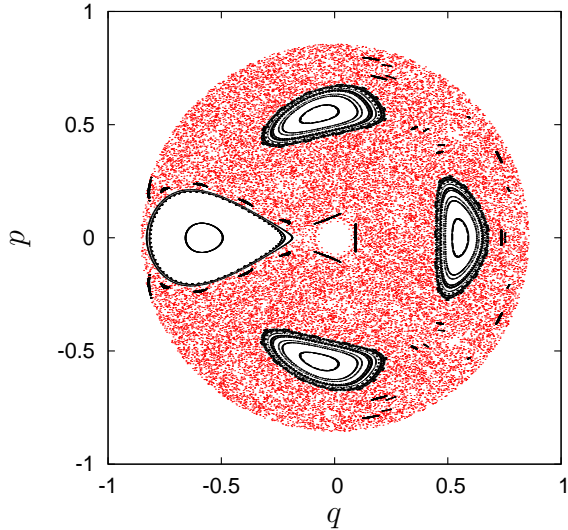}
\caption{Coordinates and momenta of the left particle
at moments of collision of the central and right ones. Red (grey) dots correspond to chaotic orbits, black dots to quasiperiodic ones.
Total energy is $E=0.8$, $E_{cm}=0$. }
\label{fig:3part}
\end{figure}

As already discussed in previous studies of the Ding-Dong model~\cite{Prosen-Robnik-92,Sano-Kitahara-01}, general 
dynamical regimes are typically chaotic. A minimal chaotic state must include three particles
 (because of the additional conservation law, for a more general non-symmetric situation chaos is possible for 
two colliding oscillators~\cite{Zheng-Su-Zhang-96}), thus two-particle regimes are periodic or quasiperiodic. 
We illustrate chaotic and quasiperiodic regimes in a 3-particle-lattice in Fig.~\ref{fig:3part}. 
Here we show a Poincar\'e map, depicting the coordinate and momentum of the left particle at the time 
instants when 
the central one and the right one collide. Remarkably, for $E_{cm}=0$ we have found that for all 
tested initial conditions the particles neighboring to the excited three ones remain untouched if the total 
energy of three initially excited particles
is less than $0.75$. This is the case of a perfectly chaotic breather that does not spread.

\section{Spreading of initially localized field}

\begin{figure}[!hbt]
\centering
\includegraphics[width=0.9\textwidth]{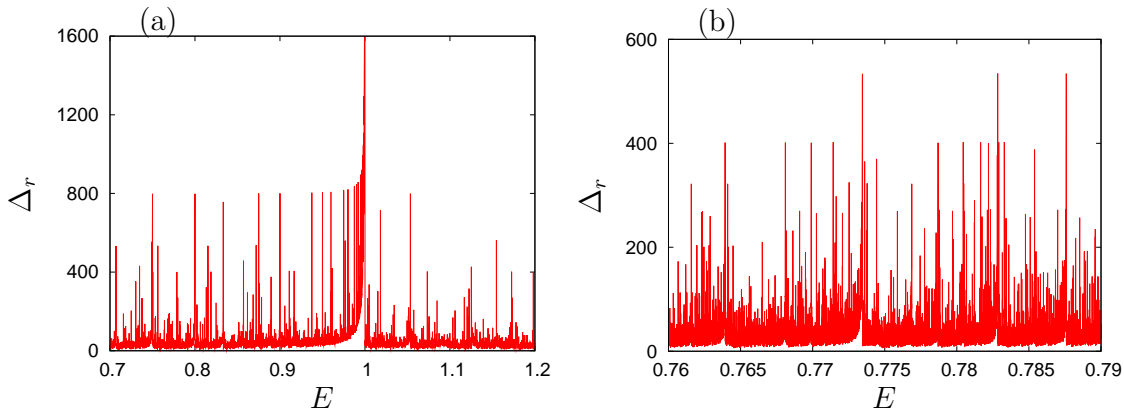}
\caption{Range of propagation from the one-site excitation, in dependence on the initial energy of the particle
$E$. Right panel is the enlarged section showing the finer structure of resonances.}
\label{fig:1part}
\end{figure}

In this section we consider properties of spreading from rather general initial conditions.
We first consider simple initial states, where the initial perturbation is restricted to one particle, 
initial energy $E$ of which is varied in a wide range. For this initial state we calculated  
the maximal range of propagation $\Delta_r$
up 
to time $t_{end}=10^4$, and plotted this range vs the initial energy in Fig.~\ref{fig:1part}. One can see ``resonances'' due to 
closeness to the compactons. The main resonance at $E=1$ in panel (a) is due to the main compacton ($n=1$) from the 
Robnik-Prosen family. There are many other resonances, as after several collisions the energy of the right-most particle
becomes close to the value for
one of the compactons, and such a perturbation propagates to a large distance. Formally, the resonances can be 
infinitely high (if perfect compactons are created) but we do not see this because of a finite resolution of initial energies.

\begin{figure}[!hbt]
\centering
\includegraphics[width=\textwidth]{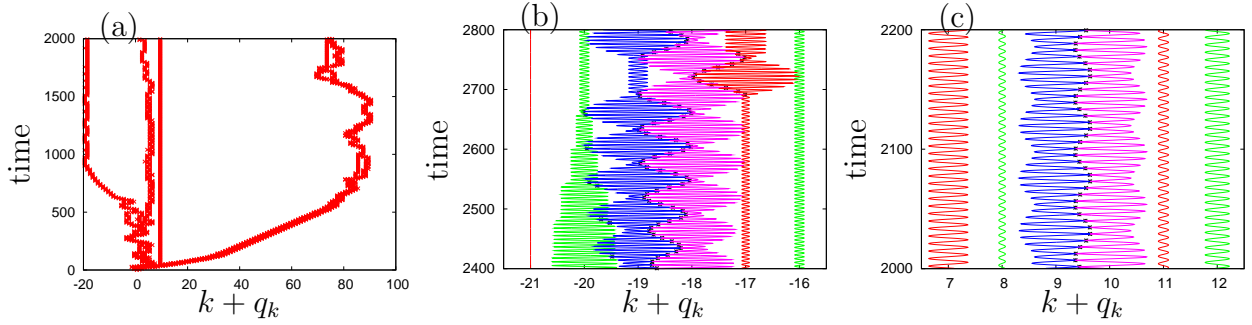}
\caption{Evolution from a particular realization of random initial momenta (5 sites are excited with total energy $E=5$). 
Panel (a) shows the pattern of collisions, in panels (b) and (c) we show collisions (with markers)
and particle trajectories (with lines) at a chaotic 
spot and at a quasiperiodic breather, respectively (see text for details).}
\label{fig:chincond}
\end{figure}

In our next setup we prepared a random initial state, setting velocities of several neighboring particles 
to random numbers. A typical evolution is shown in Fig.~\ref{fig:chincond}. Panel (a) shows the space-time pattern of 
collisions. One can see here the characteristic structures we typically observed. A compact-like wave first propagates to the right, 
close to site 30 due to losses of energy it transforms to another quasi-compacton, which propagates until site 80 where it 
gets destroyed and gives rise to an irregular collision pattern. At sites 9 and 10 a regular breather is created,
the pattern of collisions of these two particles is quasiperiodic (see panel (c)). At the left part of the space-time diagram 
one can see a chaotic quasi-breather that includes particles -20,-19, -18 and -17. Particle -21 is not excited during the 
whole time interval presented, but particles in the quasi-breather exchange energy in an erratic manner (see panel (b)).
We cannot exclude that in course of time this breather will spread or, probably, shift irregularly along the lattice performing a 
random walk. In fact, one can interpret the pattern seen at sites $70<q<90$ at times $t>500$ as 
such a randomly moving breather (``chaotic spot''). We stress here that the ``empty places'' in Fig.~\ref{fig:chincond}(a) have in fact non-zero 
energy (if they are between the left-most and the right-most sites), but this energy is insufficient for 
collisions (cf. panel (c)).

We discuss now some general properties of the spreading process, basing on the reversibility
of the dynamics. Energy cannot spread in such a way that the collisions disappear, as this would contradict
reversibility. If there has been some initial spreading stage, the final state cannot consist of quasiperiodic 
and periodic breathers
only, as such a state being reversed in time would propagate backward to infinity. Thus, the 
final state is either a perfect
combination of breathers and exact compactons, or has at least one chaotic component.  
Typically, one observes several ``active''
spots where collisions occur and which are chaotic or quasiperiodic, while the rest of the 
energy is spread along the non-active
sites that do not collide with neighbors. There is no limit for the maximal distance  in space between the right-most and 
the left-most sites, but energy cannot be uniformly distributed along the lattice because in this case collisions
would disappear. 

\begin{figure}[!hbt]
\centering
\includegraphics[width=0.9\textwidth]{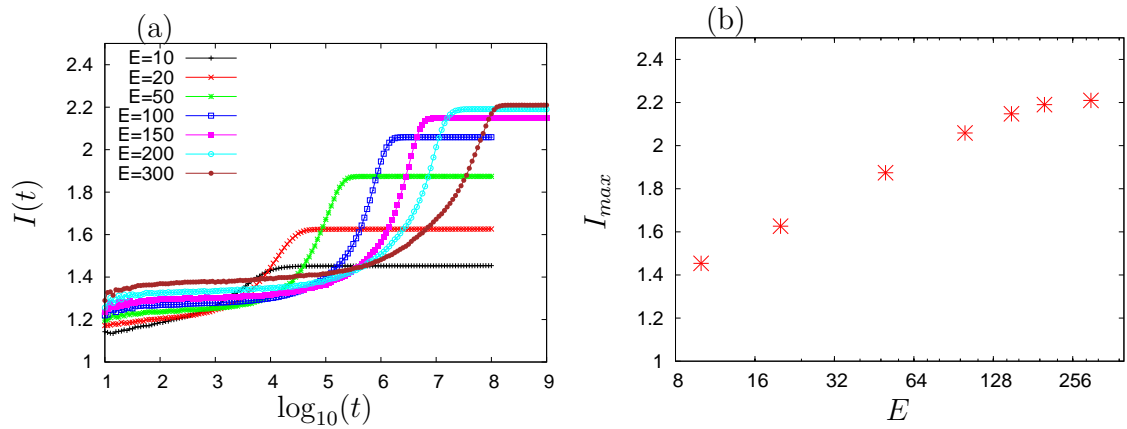}
\caption{(a): Evolution of the entropy of the energy distribution, starting from random initial conditions (momenta of particles with 
$-10\leq k\leq 10$ are random Gaussian variables), for different total energies. 
(b): Dependence of the maximal entropy on the total energy. }
\label{fig:entrevol}
\end{figure}

To characterize the spreading, we calculated the spatial entropy of the energy distribution, according to the standard
Boltzmann definition $I=-\sum_k \mathcal{E}_k\log \mathcal{E}_k$, where $\mathcal{E}_k=E_k/\hat{E}$ is the energy
at site $k$ normalized by the total energy $E$. Starting from different random initial conditions we calculated
the entropies of the configurations at different times, and than averaged them over the ensemble. The results presented
in Fig.~\ref{fig:entrevol} show that the spreading stops at some maximal entropy level, which grows with the total energy
as shown in panel (b). 

\section{Disordered Ding-Dong model}

In this section we study a disordered Ding-Dong model. Disorder can be introduced in model (\ref{eq:ddbase}) in three
ways: disorder in the spacings between the oscillators, disorder in the particle masses, and disorder in the oscillator frequencies.
The latter situation is very difficult for numerical modeling, as there is no easy way to calculate the collision time of
two harmonic oscillators with incommensurate frequencies. Thus we restrict our attention to the first two cases. In the case
of disorder in distances between the oscillators, the Hamiltonian is still  (\ref{eq:ddbase}), but the collision condition
is now $q_{k}=r_{k,k+1}+q_{k+1}$ with random spacings $ r_{k,k+1}$. At a collision, the momenta are exchanged
$p_k\to p_{k+1}$, $p_{k+1}\to p_k$. In the case of mass disorder we write the Hamiltonian as
\begin{equation}
 H=\sum_k\left(\frac{p_k^2}{2m_k}+\frac{q_k^2}{2m_k}\right)\;,
\label{eq:ddbasemd}
\end{equation}
so that all the frequencies are one. The collision condition is $q_{k}=1+q_{k+1}$, but the momenta are 
exchanged according to
\begin{equation}
  p_k\to \frac{2m_{k}p_{k+1}+(m_k-m_{k+1})p_k}{m_k+m_{k+1}}\;,\qquad
 p_{k+1}\to\frac{2m_{k+1}p_k-(m_k-m_{k+1})p_{k+1}}{m_k+m_{k+1}}\;.
\label{eq:massdis}
\end{equation}
In both cases we studied chains with the spacings or the masses independent and uniformly distributed in 
$1-\delta r<r_{k,k+1}<1+\delta r$, $1-\delta m<m_k<1+\delta m$. 

\begin{figure}[!hbt]
\centering
\includegraphics[width=0.9\textwidth]{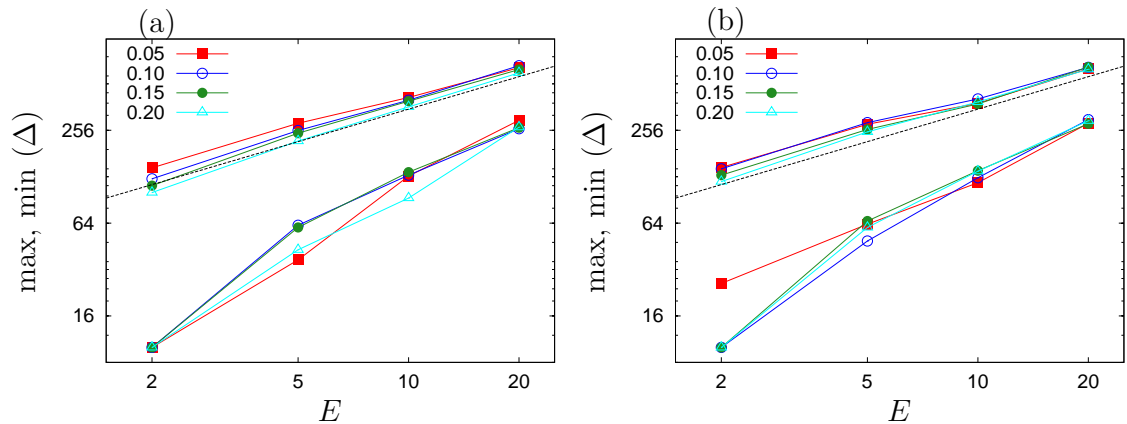}
\caption{Maximal and minimal spreading ranges $\Delta$ vs total energy,
determined from ensembles of appr. 2000 realizations of disorder and initial conditions 
(initially 10 sites were excited randomly).
(a): disorder in distances, (b): disorder in masses. Values of $\delta r$ and $\delta m$ are shown on the panels.
Dashed lines correspond to $\Delta\sim E^{0.7}$.}
\label{fig:disord}
\end{figure}

The main effect of disorder is that the compactons disappear, because the translational invariance is broken. On the other hand, 
chaotic and quasiperiodic breathers may exist.  We have determined spreading properties for different realizations of
disorder and for different initial conditions, and in all cases have found that after some initial stage the spreading eventually 
stops and
the maximal spreading range $\Delta =k_{max}-k_{min}$ remains constant. Here $k_{max}$ and $k_{min}$ are the indices of the 
right-most and the left-most excited sites, respectively. 
Of course, this conclusion is based on numerical calculations within a finite time interval. To be more precise, the conclusion
on ``non-spreading'' was made after there was no any spreading event (i.e. the left-most and the right-most excited sites 
remained unchanged) during the time interval $10^{10}$. In Fig.~\ref{fig:disord} we present the statistics of the spreading ranges for
different types of disorder, different total energies $E$, and different disorder strengths. One can see that the maximal spreading range
grows as a power law of  the total energy with power $\approx 0.7$. The spreading range 
practically does not depend on the disorder level. 

\begin{figure}[!hbt]
\centering
\includegraphics[width=\textwidth]{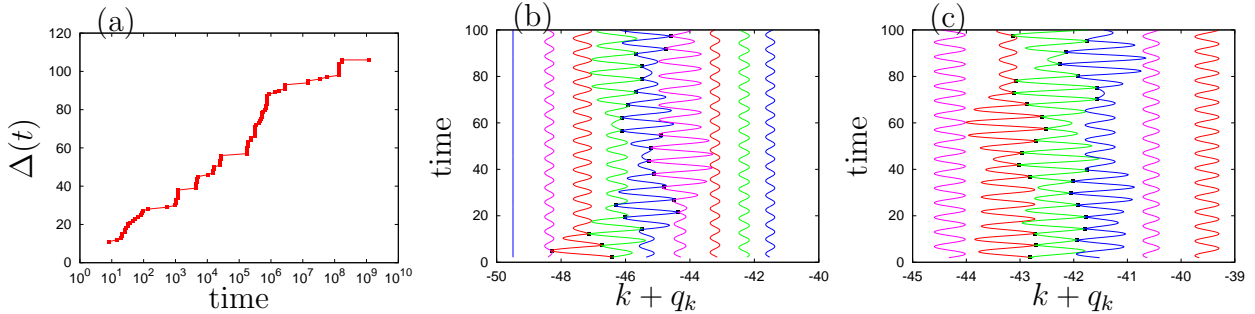}
\caption{(a): Evolution of the spreading range $\Delta$ for a particular configuration of the distance disorder, for
$\delta r=0.2$ and $E=5$. Panels (b) and (c) show a chaotic spot near the left edge of the lattice, just after the last
spreading event has happened, and after time interval $10^9$, during  which $\Delta$ remains constant,
respectively. One can see that the spot is
slightly 
shifted. Time  in (b) and (c) is marked relative to the beginning of the corresponding intervals. }
\label{fig:disord1}
\end{figure}

In some cases the initial configuration was not chaotic, in this case no spreading
was observed (thus the
minimal spreading interval is 10 for energy $E=2$), in all other cases the final state consists of a few chaotic spots. 
We illustrate this 
in Fig.~\ref{fig:disord1}. Here, for one configuration of disorder and for randomly chosen initial conditions 
we show in panel (a) how the energy spreads along the 
lattice. The maximal spreading range $\Delta=106$ is reached at $t\approx 1.7 \cdot 10^8$. 
After this time, the activity (i.e. collisions)
 is observed at two spots which very slowly shift along the lattice. 

\section{Conclusions}
In this paper we have studied properties of energy spreading in a nonlinear Ding-Dong model. This nonlinear lattice 
is not generic, as it has a strict nonlinearity threshold, below which the oscillations are purely linear and do not interact. 
We have described two
elementary excitations in the lattice - compactons and chaotic breathers. Compactons are relatively stable objects, so quite 
often we observe appearance of quasi-compactons with a long lifetime from general initial conditions. Due to this,
initially localized energy is spread over a large domain, 
but the final stage is highly nonuniform: while at many places energy is below the collision 
threshold, there exist several spots -- typically regular or chaotic breathers. Correspondingly, the spatial entropy of the energy
distribution saturates at some level depending on the total energy of the lattice. Furthermore, we studied a disordered Ding-Dong 
model. Here compactons do not exist and the energy spreading is finite: the field remains localized up to maximal times of our 
calculations. Again, several chaotic spots are formed together with a sub-threshold background.

It is instructive to compare the properties of the Ding-Dong model with those of lattices with smooth potentials. In the latter case
one also observes compactons (while with superexponential tails~\cite{Ahnert-Pikovsky-09}). 
Exact chaotic breathers appear to be impossible, 
because the interaction with the neighbors has no threshold and the neighbors would be excited by the 
noisy oscillations of such a 
breather. Similar to the Ding-Dong case, disorder destroys compactons, and the energy spreading is a slow subdiffusive process.
In a disordered smooth nonlinear lattice so far no stop of energy spreading has been observed, although there are indications that the
spreading 
slows down with the time~\cite{Johansson-Kopidakis-Aubry-10,Mulansky-Ahnert-Pikovsky-11}. Also the scaling properties 
of the largest  Lyapunov exponent in smooth lattices~\cite{Pikovsky-Fishman-11} suggest that chaos might extinct at the late stages of 
spreading. The eventual stop of spreading in the Ding-Dong model is, to the best of our knowledge, the first observation
of the localized but chaotic dynamics in nonlinear lattices. However, nonlinear disordered lattices with analytic interaction potentials possess a property that every lattice site is coupled (although indirectly) with all other sites, so that chaotic spots influence the whole lattice. This feature implicates that systems with analytic and non-analytic interaction potentials might demonstrate  different asymptotic scenaria. It would be very interesting to investigate smooth lattices close to the Ding-Dong model and to compare their localization properties  (similar to the studies of
smooth approximations of scattering billiards~\cite{Rom-Kedar-Turaev-99}).

\begin{acknowledgments}
The stay of S.R. in Potsdam was supported by DAAD. We thank M. Mulansky, R. Livi, A. Politi, M. Robnik, and 
S. Fishman for useful discussions.
\end{acknowledgments}

%


\begin{thebibliography}{10}%
\makeatletter
\providecommand \@ifxundefined [1]{%
 \ifx #1\undefined \expandafter \@firstoftwo
 \else \expandafter \@secondoftwo
\fi
}%
\providecommand \@ifnum [1]{%
 \ifnum #1\expandafter \@firstoftwo
 \else \expandafter \@secondoftwo
\fi
}%
\providecommand \enquote [1]{``#1''}%
\providecommand \bibnamefont  [1]{#1}%
\providecommand \bibfnamefont [1]{#1}%
\providecommand \citenamefont [1]{#1}%
\providecommand\href[0]{\@sanitize\@href}%
\providecommand\@href[1]{\endgroup\@@startlink{#1}\endgroup\@@href}%
\providecommand\@@href[1]{#1\@@endlink}%
\providecommand \@sanitize [0]{\begingroup\catcode`\&12\catcode`\#12\relax}%
\@ifxundefined \pdfoutput {\@firstoftwo}{%
 \@ifnum{\z@=\pdfoutput}{\@firstoftwo}{\@secondoftwo}%
}{%
 \providecommand\@@startlink[1]{\leavevmode}%
 \providecommand\@@endlink[0]{}%
}{%
 \providecommand\@@startlink[1]{%
  \leavevmode
  \pdfstartlink
   attr{/Border[0 0 1 ]/H/I/C[0 1 1]}%
   user{/Subtype/Link/A<</Type/Action/S/URI/URI(#1)>>}%
  \relax
 }%
 \providecommand\@@endlink[0]{\pdfendlink}%
}%
\providecommand \url  [0]{\begingroup\@sanitize \@url }%
\providecommand \@url [1]{\endgroup\@href {#1}{\urlprefix}}%
\providecommand \urlprefix [0]{URL }%
\providecommand \Eprint[0]{\href }%
\@ifxundefined \urlstyle {%
  \providecommand \doi [1]{doi:\discretionary{}{}{}#1}%
}{%
  \providecommand \doi [0]{doi:\discretionary{}{}{}\begingroup
  \urlstyle{rm}\Url }%
}%
\providecommand \doibase [0]{http://dx.doi.org/}%
\providecommand \Doi[1]{\href{\doibase#1}}%
\providecommand \selectlanguage [0]{\@gobble}%
\providecommand \bibinfo [0]{\@secondoftwo}%
\providecommand \bibfield [0]{\@secondoftwo}%
\providecommand \translation [1]{[#1]}%
\providecommand \BibitemOpen[0]{}%
\providecommand \bibitemStop [0]{}%
\providecommand \bibitemNoStop [0]{.\EOS\space}%
\providecommand \EOS [0]{\spacefactor3000\relax}%
\providecommand \BibitemShut [1]{\csname bibitem#1\endcsname}%
\bibitem{Gallavotti-08}%
  \BibitemOpen
  \emph{\bibinfo {title} {The Fermi-Pasta-Ulam problem}},\ edited by\ \bibinfo
  {editor} {\bibfnamefont{G.}~\bibnamefont{Gallavotti}}\ (\bibinfo {publisher}
  {Springer Lecture Notes in Physics vol. 728},\ \bibinfo {year}
  {2008})\BibitemShut{NoStop}%
\bibitem{Chaos-fpu-05}%
  \BibitemOpen
  \bibfield{author}{%
  \bibinfo {author} {\bibnamefont{{A focus issue on ``The ``Fermi-Pasta-Ulam"
  problem -- the first 50 years'' (ed. by D. K. Campbell, P. Rosenau and G.
  Zaslavsky)}}},\ }%
  \bibfield{journal}{%
  \bibinfo {journal} {CHAOS}\ }%
  \textbf{\bibinfo {volume} {15}} (\bibinfo {year} {2005})\BibitemShut{NoStop}%
\bibitem{Sinai-Chernov-87}%
  \BibitemOpen
  \bibfield{author}{%
  \bibinfo {author} {\bibfnamefont{Ya.~G.}\ \bibnamefont{Sinai}}\ and\ \bibinfo
  {author} {\bibfnamefont{N.~I.}\ \bibnamefont{Chernov}},\ }%
  \bibfield{title}{%
  \enquote{\bibinfo {title} {Ergodic properties of certain systems of
  two-dimensional discs and three-dimensional balls},}\ }%
  \bibfield{journal}{%
  \bibinfo {journal} {Russian Math. Surveys}\ }%
  \textbf{\bibinfo {volume} {42}},\ \bibinfo {pages} {181--207} (\bibinfo
  {year} {1987})\BibitemShut{NoStop}%
\bibitem{Casati84}%
  \BibitemOpen
  \bibfield{author}{%
  \bibinfo {author} {\bibfnamefont{G.}~\bibnamefont{Casati}}, \bibinfo {author}
  {\bibfnamefont{J.}~\bibnamefont{Ford}}, \bibinfo {author}
  {\bibfnamefont{F.}~\bibnamefont{Vivaldi}},\ and\ \bibinfo {author}
  {\bibfnamefont{W.~M.}\ \bibnamefont{Visscher}},\ }%
  \bibfield{title}{%
  \enquote{\bibinfo {title} {One-dimensional classical many-body system having
  a normal thermal conductivity},}\ }%
  \bibfield{journal}{%
  \bibinfo {journal} {Phys. Rev. Lett.}\ }%
  \textbf{\bibinfo {volume} {52}},\ \bibinfo {pages} {1961} (\bibinfo {year}
  {1984})\BibitemShut{NoStop}%
\bibitem{Casati-86}%
  \BibitemOpen
  \bibfield{author}{%
  \bibinfo {author} {\bibfnamefont{G.}~\bibnamefont{Casati}},\ }%
  \bibfield{title}{%
  \enquote{\bibinfo {title} {Energy transport and the {F}ourier law in
  classical systems},}\ }%
  \bibfield{journal}{%
  \bibinfo {journal} {Found. Physics}\ }%
  \textbf{\bibinfo {volume} {16}},\ \bibinfo {pages} {51--61} (\bibinfo {year}
  {1985})\BibitemShut{NoStop}%
\bibitem{Posch-Hoover-98}%
  \BibitemOpen
  \bibfield{author}{%
  \bibinfo {author} {\bibfnamefont{H.~A.}\ \bibnamefont{Posch}}\ and\ \bibinfo
  {author} {\bibfnamefont{Wm.~G.}\ \bibnamefont{Hoover}},\ }%
  \bibfield{title}{%
  \enquote{\bibinfo {title} {Heat conduction in one-dimensional chains and
  nonequilibrium {L}yapunov spectrum},}\ }%
  \bibfield{journal}{%
  \bibinfo {journal} {Phys. Rev. E}\ }%
  \textbf{\bibinfo {volume} {58}},\ \bibinfo {pages} {4344--4350} (\bibinfo
  {month} {Oct}\ \bibinfo {year} {1998})\BibitemShut{NoStop}%
\bibitem{Gawronski-Kulakovski-02}%
  \BibitemOpen
  \bibfield{author}{%
  \bibinfo {author} {\bibfnamefont{P.}~\bibnamefont{Gawronski}}\ and\ \bibinfo
  {author} {\bibfnamefont{K.}~\bibnamefont{Kulakovski}},\ }%
  \bibfield{title}{%
  \enquote{\bibinfo {title} {Chaos of two particles in the ding-a-ling
  model},}\ }%
  \bibfield{journal}{%
  \bibinfo {journal} {Computer Physics Communications}\ }%
  \textbf{\bibinfo {volume} {147}},\ \bibinfo {pages} {608--611} (\bibinfo
  {year} {2002})\BibitemShut{NoStop}%
\bibitem{Prosen-Robnik-92}%
  \BibitemOpen
  \bibfield{author}{%
  \bibinfo {author} {\bibfnamefont{T.}~\bibnamefont{Prosen}}\ and\ \bibinfo
  {author} {\bibfnamefont{M.}~\bibnamefont{Robnik}},\ }%
  \bibfield{title}{%
  \enquote{\bibinfo {title} {Energy transport and detailed verification of
  fourier law in a chain of colliding harmonic oscillators},}\ }%
  \bibfield{journal}{%
  \bibinfo {journal} {J. Phys. A}\ }%
  \textbf{\bibinfo {volume} {25}},\ \bibinfo {pages} {3449--3472} (\bibinfo
  {year} {1992})\BibitemShut{NoStop}%
\bibitem{Shepelyansky-93}%
  \BibitemOpen
  \bibfield{author}{%
  \bibinfo {author} {\bibfnamefont{D.}~\bibnamefont{Shepelyansky}},\ }%
  \bibfield{journal}{%
  \bibinfo {journal} {Phys. Rev. Lett.}}%
   (\bibinfo {year} {1993})\BibitemShut{NoStop}%
\bibitem{Molina-98}%
  \BibitemOpen
  \bibfield{author}{%
  \bibinfo {author} {\bibfnamefont{M.~I.}\ \bibnamefont{Molina}},\ }%
  \bibfield{title}{%
  \enquote{\bibinfo {title} {Transport of localized and extended excitations in
  a nonlinear {A}nderson model},}\ }%
  \bibfield{journal}{%
  \bibinfo {journal} {Phys. Rev. B}\ }%
  \textbf{\bibinfo {volume} {58}},\ \bibinfo {pages} {12547--12550} (\bibinfo
  {year} {1998})\BibitemShut{NoStop}%
\bibitem{Pikovsky-Shepelyansky-08}%
  \BibitemOpen
  \bibfield{author}{%
  \bibinfo {author} {\bibfnamefont{A.~S.}\ \bibnamefont{Pikovsky}}\ and\
  \bibinfo {author} {\bibfnamefont{D.~L.}\ \bibnamefont{Shepelyansky}},\ }%
  \bibfield{title}{%
  \enquote{\bibinfo {title} {Destruction of {A}nderson localization by a weak
  nonlinearity},}\ }%
  \bibfield{journal}{%
  \bibinfo {journal} {Phys. Rev. Lett.}\ }%
  \textbf{\bibinfo {volume} {100}},\ \bibinfo {pages} {094101} (\bibinfo {year}
  {2008})\BibitemShut{NoStop}%
\bibitem{Garcia-Mata-Shepelyansky-09}%
  \BibitemOpen
  \bibfield{author}{%
  \bibinfo {author} {\bibfnamefont{I.}~\bibnamefont{Garcia-Mata}}\ and\
  \bibinfo {author} {\bibfnamefont{D.~L.}\ \bibnamefont{Shepelyansky}},\ }%
  \bibfield{title}{%
  \enquote{\bibinfo {title} {{Nonlinear delocalization on disordered {S}tark
  ladder}},}\ }%
  \bibfield{journal}{%
  \bibinfo {journal} {{Eur. Phys. J. B}}\ }%
  \textbf{\bibinfo {volume} {{71}}},\ \bibinfo {pages} {{121--124}} (\bibinfo
  {year} {{2009}})\BibitemShut{NoStop}%
\bibitem{Flach-Krimer-Skokos-09}%
  \BibitemOpen
  \bibfield{author}{%
  \bibinfo {author} {\bibfnamefont{S.}~\bibnamefont{Flach}}, \bibinfo {author}
  {\bibfnamefont{D.~O.}\ \bibnamefont{Krimer}},\ and\ \bibinfo {author}
  {\bibfnamefont{Ch.}\ \bibnamefont{Skokos}},\ }%
  \bibfield{title}{%
  \enquote{\bibinfo {title} {Universal spreading of wave packets in disordered
  nonlinear systems},}\ }%
  \bibfield{journal}{%
  \bibinfo {journal} {Phys. Rev. Lett.}\ }%
  \textbf{\bibinfo {volume} {102}},\ \bibinfo {pages} {024101} (\bibinfo {year}
  {2009})\BibitemShut{NoStop}%
\bibitem{Skokos_etal-09}%
  \BibitemOpen
  \bibfield{author}{%
  \bibinfo {author} {\bibfnamefont{Ch.}\ \bibnamefont{Skokos}}, \bibinfo
  {author} {\bibfnamefont{D.~O.}\ \bibnamefont{Krimer}}, \bibinfo {author}
  {\bibfnamefont{S.}~\bibnamefont{Komineas}},\ and\ \bibinfo {author}
  {\bibfnamefont{S.}~\bibnamefont{Flach}},\ }%
  \bibfield{title}{%
  \enquote{\bibinfo {title} {{Delocalization of wave packets in disordered
  nonlinear chains}},}\ }%
  \bibfield{journal}{%
  \bibinfo {journal} {{Phys. Rev. E}}\ }%
  \textbf{\bibinfo {volume} {{79}}},\ \bibinfo {pages} {{056211}} (\bibinfo
  {year} {{2009}})\BibitemShut{NoStop}%
\bibitem{Mulansky-Ahnert-Pikovsky-Shepelyansky-09}%
  \BibitemOpen
  \bibfield{author}{%
  \bibinfo {author} {\bibfnamefont{M.}~\bibnamefont{Mulansky}}, \bibinfo
  {author} {\bibfnamefont{K.}~\bibnamefont{Ahnert}}, \bibinfo {author}
  {\bibfnamefont{A.}~\bibnamefont{Pikovsky}},\ and\ \bibinfo {author}
  {\bibfnamefont{D.~L.}\ \bibnamefont{Shepelyansky}},\ }%
  \bibfield{title}{%
  \enquote{\bibinfo {title} {Dynamical thermalization of disordered nonlinear
  lattices},}\ }%
  \bibfield{journal}{%
  \bibinfo {journal} {Phys. Rev. E}\ }%
  \textbf{\bibinfo {volume} {80}},\ \bibinfo {pages} {056212} (\bibinfo {year}
  {2009})\BibitemShut{NoStop}%
\bibitem{Skokos-Flach-10}%
  \BibitemOpen
  \bibfield{author}{%
  \bibinfo {author} {\bibnamefont{Ch.Skokos}}\ and\ \bibinfo {author}
  {\bibfnamefont{S.}~\bibnamefont{Flach}},\ }%
  \bibfield{title}{%
  \enquote{\bibinfo {title} {Spreading of wave packets in disordered systems
  with tunable nonlinearity},}\ }%
  \bibfield{journal}{%
  \bibinfo {journal} {Phys. Rev. E}\ }%
  \textbf{\bibinfo {volume} {82}},\ \bibinfo {pages} {016208} (\bibinfo {year}
  {2010})\BibitemShut{NoStop}%
\bibitem{Flach-10}%
  \BibitemOpen
  \bibfield{author}{%
  \bibinfo {author} {\bibfnamefont{S.}~\bibnamefont{Flach}},\ }%
  \bibfield{title}{%
  \enquote{\bibinfo {title} {{Spreading of waves in nonlinear disordered
  media}},}\ }%
  \bibfield{journal}{%
  \bibinfo {journal} {{Chem. Physics}}\ }%
  \textbf{\bibinfo {volume} {{375}}},\ \bibinfo {pages} {{548--556}} (\bibinfo
  {month} {{OCT 5}}\ \bibinfo {year} {{2010}})\BibitemShut{NoStop}%
\bibitem{Laptyeva-etal-10}%
  \BibitemOpen
  \bibfield{author}{%
  \bibinfo {author} {\bibfnamefont{T.~V.}\ \bibnamefont{Laptyeva}}, \bibinfo
  {author} {\bibfnamefont{J.~D.}\ \bibnamefont{Bodyfelt}}, \bibinfo {author}
  {\bibfnamefont{D.~O.}\ \bibnamefont{Krimer}}, \bibinfo {author}
  {\bibnamefont{Ch.Skokos}},\ and\ \bibinfo {author}
  {\bibfnamefont{S.}~\bibnamefont{Flach}},\ }%
  \bibfield{title}{%
  \enquote{\bibinfo {title} {The crossover from strong to weak chaos for
  nonlinear waves in disordered systems},}\ }%
  \bibfield{journal}{%
  \bibinfo {journal} {Europhys. Lett.}\ }%
  \textbf{\bibinfo {volume} {91}},\ \bibinfo {pages} {30001} (\bibinfo {year}
  {2010})\BibitemShut{NoStop}%
\bibitem{Mulansky-Pikovsky-10}%
  \BibitemOpen
  \bibfield{author}{%
  \bibinfo {author} {\bibfnamefont{M.}~\bibnamefont{Mulansky}}\ and\ \bibinfo
  {author} {\bibfnamefont{A.}~\bibnamefont{Pikovsky}},\ }%
  \bibfield{title}{%
  \enquote{\bibinfo {title} {Spreading in disordered lattices with different
  nonlinearities},}\ }%
  \bibfield{journal}{%
  \bibinfo {journal} {Europhys. Lett.}\ }%
  \textbf{\bibinfo {volume} {{90}}},\ \bibinfo {pages} {{10015}} (\bibinfo
  {year} {2010})\BibitemShut{NoStop}%
\bibitem{Johansson-Kopidakis-Aubry-10}%
  \BibitemOpen
  \bibfield{author}{%
  \bibinfo {author} {\bibfnamefont{M.}~\bibnamefont{Johansson}}, \bibinfo
  {author} {\bibfnamefont{G.}~\bibnamefont{Kopidakis}},\ and\ \bibinfo {author}
  {\bibfnamefont{S.}~\bibnamefont{Aubry}},\ }%
  \bibfield{title}{%
  \enquote{\bibinfo {title} {{KAM} tori in {1D} random discrete nonlinear
  {S}chr{\"o}dinger model?}.}\ }%
  \bibfield{journal}{%
  \bibinfo {journal} {Europhys. Lett.}\ }%
  \textbf{\bibinfo {volume} {91}},\ \bibinfo {pages} {50001} (\bibinfo {year}
  {2010})\BibitemShut{NoStop}%
\bibitem{Mulansky-Ahnert-Pikovsky-11}%
  \BibitemOpen
  \bibfield{author}{%
  \bibinfo {author} {\bibfnamefont{M.}~\bibnamefont{Mulansky}}, \bibinfo
  {author} {\bibfnamefont{K.}~\bibnamefont{Ahnert}},\ and\ \bibinfo {author}
  {\bibfnamefont{A.}~\bibnamefont{Pikovsky}},\ }%
  \bibfield{title}{%
  \enquote{\bibinfo {title} {Scaling of energy spreading in strongly nonlinear
  disordered lattices},}\ }%
  \bibfield{journal}{%
  \bibinfo {journal} {Phys. Rev. E}\ }%
  \textbf{\bibinfo {volume} {83}},\ \bibinfo {pages} {026205} (\bibinfo {year}
  {2011})\BibitemShut{NoStop}%
\bibitem{Sano-00}%
  \BibitemOpen
  \bibfield{author}{%
  \bibinfo {author} {\bibfnamefont{M.~M.}\ \bibnamefont{Sano}},\ }%
  \bibfield{title}{%
  \enquote{\bibinfo {title} {Equilibrium and stationary nonequilibrium states
  in a chain of colliding harmonic oscillators},}\ }%
  \bibfield{journal}{%
  \bibinfo {journal} {Phys. Rev. E}\ }%
  \textbf{\bibinfo {volume} {61}},\ \bibinfo {pages} {1144--1151} (\bibinfo
  {year} {2000})\BibitemShut{NoStop}%
\bibitem{Sano-Kitahara-01}%
  \BibitemOpen
  \bibfield{author}{%
  \bibinfo {author} {\bibfnamefont{M.~M.}\ \bibnamefont{Sano}}\ and\ \bibinfo
  {author} {\bibfnamefont{K.}~\bibnamefont{Kitahara}},\ }%
  \bibfield{title}{%
  \enquote{\bibinfo {title} {Thermal conduction in a chain of colliding
  harmonic oscillators revisited},}\ }%
  \bibfield{journal}{%
  \bibinfo {journal} {Phys. Rev. E}\ }%
  \textbf{\bibinfo {volume} {64}},\ \bibinfo {pages} {056111} (\bibinfo {year}
  {2001})\BibitemShut{NoStop}%
\bibitem{Sano-06}%
  \BibitemOpen
  \bibfield{author}{%
  \bibinfo {author} {\bibfnamefont{M.~M.}\ \bibnamefont{Sano}},\ }%
  \bibfield{title}{%
  \enquote{\bibinfo {title} {Steady thermal conduction far from equilibrium in
  {D}ing-{D}ong model},}\ }%
  \bibfield{journal}{%
  \bibinfo {journal} {J. Phys. Soc. Japan}\ }%
  \textbf{\bibinfo {volume} {75}},\ \bibinfo {pages} {094002} (\bibinfo {year}
  {2006})\BibitemShut{NoStop}%
\bibitem{Rosenau-Hyman-93}%
  \BibitemOpen
  \bibfield{author}{%
  \bibinfo {author} {\bibfnamefont{P.}~\bibnamefont{Rosenau}}\ and\ \bibinfo
  {author} {\bibfnamefont{J.~M.}\ \bibnamefont{Hyman}},\ }%
  \bibfield{title}{%
  \enquote{\bibinfo {title} {Compactons: {S}olitons with finite wavelength},}\
  }%
  \bibfield{journal}{%
  \bibinfo {journal} {Phys. Rev. Lett.}\ }%
  \textbf{\bibinfo {volume} {70}},\ \bibinfo {pages} {564--567} (\bibinfo
  {year} {1993})\BibitemShut{NoStop}%
\bibitem{Rosenau-94}%
  \BibitemOpen
  \bibfield{author}{%
  \bibinfo {author} {\bibfnamefont{P.}~\bibnamefont{Rosenau}},\ }%
  \bibfield{title}{%
  \enquote{\bibinfo {title} {Nonlinear dispersion and compact structures},}\ }%
  \bibfield{journal}{%
  \bibinfo {journal} {Phys. Rev. Lett.}\ }%
  \textbf{\bibinfo {volume} {73}},\ \bibinfo {pages} {1737--1741} (\bibinfo
  {year} {1994})\BibitemShut{NoStop}%
\bibitem{Rosenau-06}%
  \BibitemOpen
  \bibfield{author}{%
  \bibinfo {author} {\bibfnamefont{P.}~\bibnamefont{Rosenau}},\ }%
  \bibfield{title}{%
  \enquote{\bibinfo {title} {On a model equation of traveling and stationary
  compactons},}\ }%
  \bibfield{journal}{%
  \bibinfo {journal} {Phys. Lett. A}\ }%
  \textbf{\bibinfo {volume} {356}},\ \bibinfo {pages} {44} (\bibinfo {year}
  {2006})\BibitemShut{NoStop}%
\bibitem{Rosenau-Schochet-07}%
  \BibitemOpen
  \bibfield{author}{%
  \bibinfo {author} {\bibfnamefont{P.}~\bibnamefont{Rosenau}}\ and\ \bibinfo
  {author} {\bibfnamefont{S.}~\bibnamefont{Schochet}},\ }%
  \bibfield{title}{%
  \enquote{\bibinfo {title} {Almost compact breathers in anharmonic lattices
  near the continuum limit},}\ }%
  \bibfield{journal}{%
  \bibinfo {journal} {Phys. Rev. Lett.}\ }%
  \textbf{\bibinfo {volume} {94}},\ \bibinfo {pages} {045503} (\bibinfo {year}
  {2005})\BibitemShut{NoStop}%
\bibitem{Rosenau-Pikovsky-05}%
  \BibitemOpen
  \bibfield{author}{%
  \bibinfo {author} {\bibfnamefont{P.}~\bibnamefont{Rosenau}}\ and\ \bibinfo
  {author} {\bibfnamefont{A.}~\bibnamefont{Pikovsky}},\ }%
  \bibfield{title}{%
  \enquote{\bibinfo {title} {Phase compactons in chains of dispersively coupled
  oscillators},}\ }%
  \bibfield{journal}{%
  \bibinfo {journal} {Phys. Rev. Lett.}\ }%
  \textbf{\bibinfo {volume} {94}},\ \bibinfo {pages} {174102} (\bibinfo {year}
  {2005})\BibitemShut{NoStop}%
\bibitem{Pikovsky-Rosenau-06}%
  \BibitemOpen
  \bibfield{author}{%
  \bibinfo {author} {\bibfnamefont{A.}~\bibnamefont{Pikovsky}}\ and\ \bibinfo
  {author} {\bibfnamefont{P.}~\bibnamefont{Rosenau}},\ }%
  \bibfield{title}{%
  \enquote{\bibinfo {title} {Phase compactons},}\ }%
  \bibfield{journal}{%
  \bibinfo {journal} {Physica D}\ }%
  \textbf{\bibinfo {volume} {218}},\ \bibinfo {pages} {56--69} (\bibinfo {year}
  {2006})\BibitemShut{NoStop}%
\bibitem{Ahnert-Pikovsky-08}%
  \BibitemOpen
  \bibfield{author}{%
  \bibinfo {author} {\bibfnamefont{K.}~\bibnamefont{Ahnert}}\ and\ \bibinfo
  {author} {\bibfnamefont{A.}~\bibnamefont{Pikovsky}},\ }%
  \bibfield{title}{%
  \enquote{\bibinfo {title} {Traveling waves and compactons in phase oscillator
  lattices},}\ }%
  \bibfield{journal}{%
  \bibinfo {journal} {CHAOS}\ }%
  \textbf{\bibinfo {volume} {18}},\ \bibinfo {pages} {037118} (\bibinfo {year}
  {2008})\BibitemShut{NoStop}%
\bibitem{Ahnert-Pikovsky-09}%
  \BibitemOpen
  \bibfield{author}{%
  \bibinfo {author} {\bibfnamefont{K.}~\bibnamefont{Ahnert}}\ and\ \bibinfo
  {author} {\bibfnamefont{A.}~\bibnamefont{Pikovsky}},\ }%
  \bibfield{title}{%
  \enquote{\bibinfo {title} {Compactons and chaos in strongly nonlinear
  lattices},}\ }%
  \bibfield{journal}{%
  \bibinfo {journal} {Phys. Rev. E}\ }%
  \textbf{\bibinfo {volume} {79}},\ \bibinfo {pages} {026209} (\bibinfo {year}
  {2009})\BibitemShut{NoStop}%
\bibitem{Zheng-Su-Zhang-96}%
  \BibitemOpen
  \bibfield{author}{%
  \bibinfo {author} {\bibfnamefont{Q.-R.}\ \bibnamefont{Zheng}}, \bibinfo
  {author} {\bibfnamefont{G.}~\bibnamefont{Su}},\ and\ \bibinfo {author}
  {\bibfnamefont{D.-H.}\ \bibnamefont{Zhang}},\ }%
  \bibfield{title}{%
  \enquote{\bibinfo {title} {Classical and quantum chaotic behaviors of two
  colliding harmonic oscillators},}\ }%
  \bibfield{journal}{%
  \bibinfo {journal} {Phys. Lett. A}\ }%
  \textbf{\bibinfo {volume} {212}},\ \bibinfo {pages} {138--148} (\bibinfo
  {year} {1996})\BibitemShut{NoStop}%
\bibitem{Pikovsky-Fishman-11}%
  \BibitemOpen
  \bibfield{author}{%
  \bibinfo {author} {\bibfnamefont{A.}~\bibnamefont{Pikovsky}}\ and\ \bibinfo
  {author} {\bibfnamefont{S.}~\bibnamefont{Fishman}},\ }%
  \bibfield{title}{%
  \enquote{\bibinfo {title} {Scaling properties of weak chaos in nonlinear
  disordered lattices},}\ }%
  \bibfield{journal}{%
  \bibinfo {journal} {Phys. Rev. E}\ }%
  \textbf{\bibinfo {volume} {83}},\ \bibinfo {pages} {025201(R)} (\bibinfo
  {year} {2011})\BibitemShut{NoStop}%
\bibitem{Rom-Kedar-Turaev-99}%
  \BibitemOpen
  \bibfield{author}{%
  \bibinfo {author} {\bibfnamefont{V.}~\bibnamefont{Rom-Kedar}}\ and\ \bibinfo
  {author} {\bibfnamefont{D.}~\bibnamefont{Turaev}},\ }%
  \bibfield{title}{%
  \enquote{\bibinfo {title} {Big islands in dispersing billiard-like potentials},}\ }%
  \bibfield{journal}{%
  \bibinfo {journal} {Physica D}\ }%
  \textbf{\bibinfo {volume} {130}},\ \bibinfo {pages} {187} (\bibinfo
  {year} {1999})\BibitemShut{NoStop}%
\end{thebibliography}
\end{document}